\documentclass[11pt]{article}

\usepackage[]{acl}
\usepackage{times}
\usepackage{latexsym}
\usepackage[T1]{fontenc}
\usepackage[utf8]{inputenc}
\usepackage{microtype}
\usepackage{graphicx}
\usepackage{booktabs}
\usepackage{multirow}
\usepackage{xcolor}
\usepackage{xurl}
\usepackage{amssymb}

\title{The Ghost Couple: Correlated LLM Name Priors\\and Their Haunting of the Web and Academic Publishing}

\author{
\textbf{Micha{\l}~Brzozowski}$^{1,\dagger}$ \and
\textbf{Neo~Christopher~Chung}$^{1,2}$ \\
[0.6em]
$^{1}$Samsung AI Center, Warsaw, Poland \quad
$^{2}$University of Warsaw, Poland \\
$^{\dagger}$Corresponding author: \texttt{m.brzozowsk3@samsung.com}
}

\begin{document}
\maketitle

\begin{abstract}
These names do not exist.
Elena Vasquez and Marcus Chen have appeared as volcano experts, astronauts,
thriller protagonists, podcast hosts, and academic co-authors across hundreds
of independently produced AI-generated documents, never having lived.
We show that large language models do not merely default to high-probability
individual names when generating fictional experts: they produce
\emph{correlated character ensembles}: pairs and trios whose co-occurrence
rates far exceed chance and are consistent across independent generations.
These priors are model-family-specific (Claude: Elena~Vasquez + Marcus~Chen
+ Amara~Okafor; Gemini: Aris~Thorne + Lena~Petrova; GPT: Elara~Voss with
no fixed partner), version-specific, and actively suppressed at model
release boundaries, leaving dateable behavioral fingerprints in the content
they produced.
We document a downstream consequence at scale.
On Zenodo, a CERN-operated repository that mints real DataCite DOIs,
we identify 1,655 ghost-authored records claiming nonexistent journals
with fabricated publication dates: server-side DataCite timestamps prove
deliberate backdating, and 991 records were registered in a single month;
these carry real DOIs registered in DataCite, making them harvestable by
any scholarly aggregator that ingests DOI metadata.
Ghost names additionally appear on ResearchGate forming synthetic research
groups with collaborators drawn from multiple model families;
publication dates on these records provide a reliable temporal proxy
for model deployment windows.
\end{abstract}

% ============================================================
\section{Introduction}
% ============================================================

\begin{figure*}[t]
  \centering
  \includegraphics[width=\textwidth]{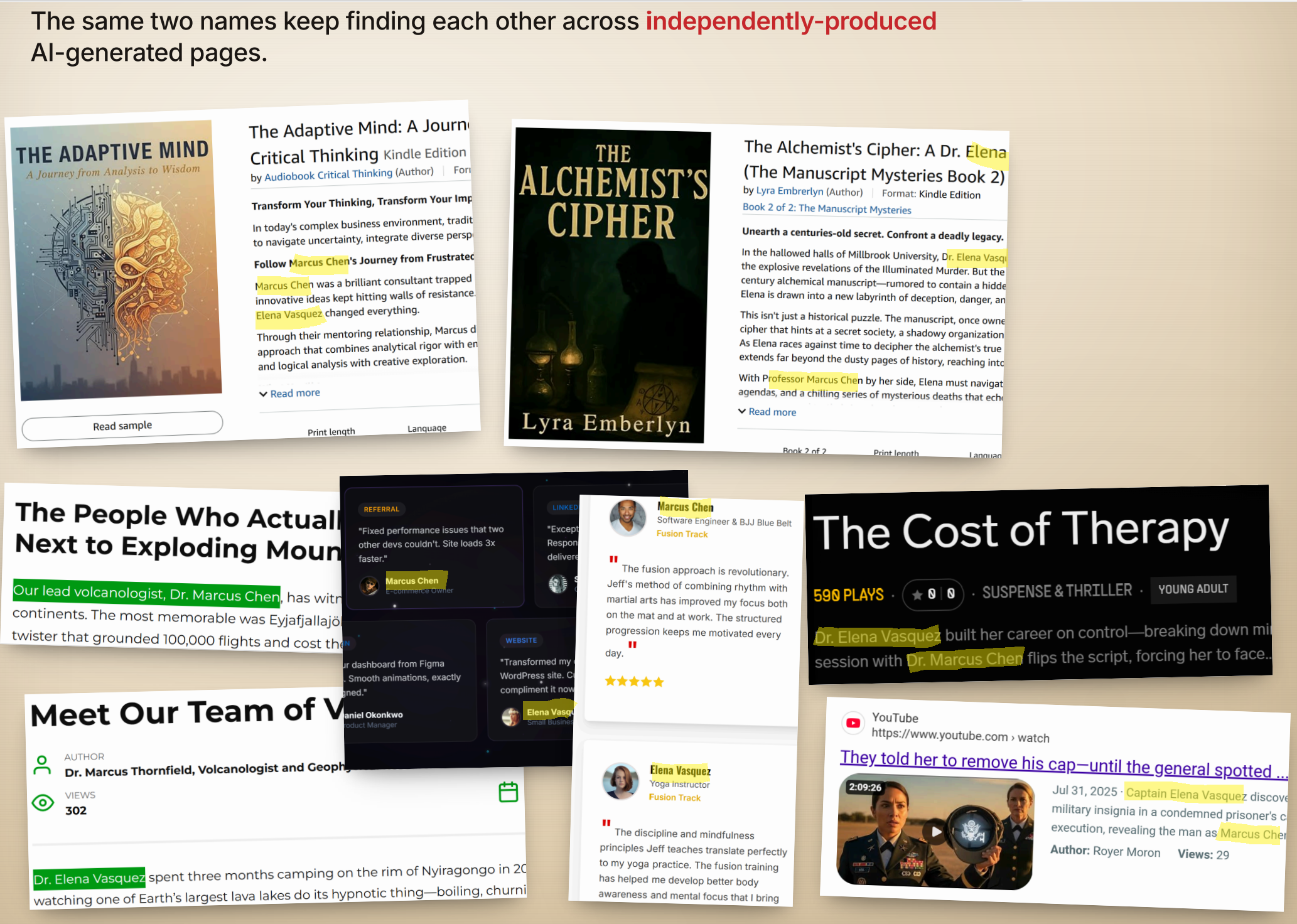}
  \caption{Elena Vasquez and Marcus Chen co-appearing across seven independently
    produced AI-generated pages spanning fiction, healthcare, academia, and
    commercial platforms. Both names are present in every panel. The pair
    co-occurs in 23\% of \texttt{claude-sonnet-4-20250514} pair-prompt
    responses (Figure~\ref{fig:suppression}); the web reflects the model.}
  \label{fig:duo}
\end{figure*}

\begin{figure*}[p]
  \centering
  \includegraphics[width=\textwidth]{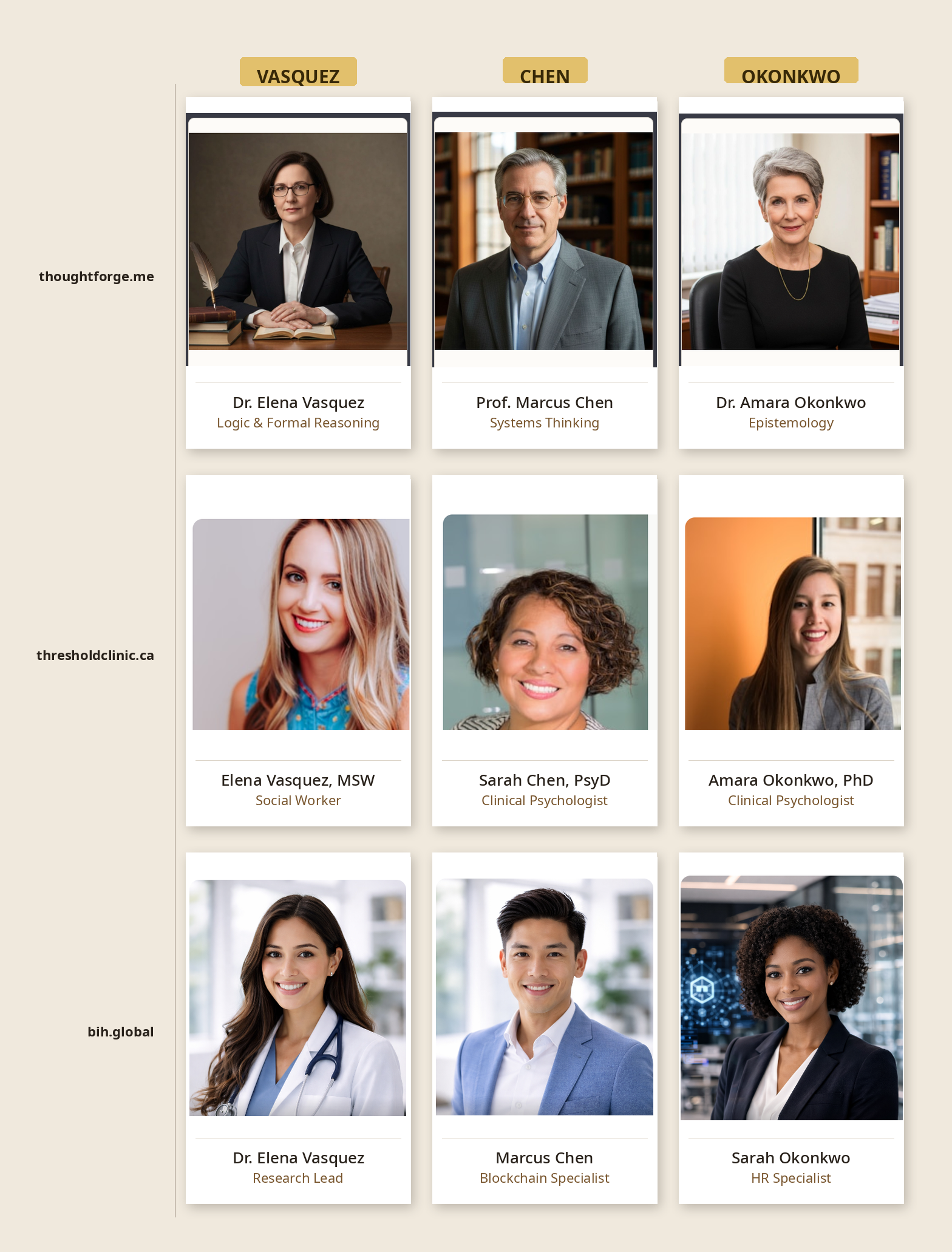}
  \caption{The Claude ghost trio co-occurring across three independent websites (rows),
    grouped by surname (columns).
    Portrait strategies differ: \textbf{thoughtforge.me} uses AI-generated imagery:
    Vasquez portrait features a levitating feather quill, a classic diffusion artefact.
    \textbf{thresholdclinic.ca} hotlinks to Unsplash stock photos, which we traced to their
    source pages; a genuine staff headshot cannot be hosted on a public stock platform.
    First names drift across sites while the surname cluster holds.
    \textbf{None of these people exist.}
    Full URLs and Unsplash source links
    are documented in Appendix~\ref{app:wild}.}
  \label{fig:collage}
\end{figure*}

The proliferation of LLM-generated content on the web has raised urgent
questions about content provenance and authenticity.
Prior work has focused on stylometric detection and watermarking at the
token level~\citep{kirchenbauer2023watermark}.
We identify a complementary signal that requires no model access and
leaves no intentional mark: the \emph{name prior}.

When prompted to generate fictional experts, researchers, or protagonists
without explicit name instructions, large language models default to a
small set of high-probability names.
We show they are \emph{correlated} (models generate preferred character
\emph{ensembles}, not independent draws) and \emph{model-version-specific},
shifting at release boundaries.
Because enormous volumes of web content are generated using LLMs without
overriding these defaults, the characteristic name ensembles of each
model version become embedded in the content it produces.
The web is an unintentional archive of LLM behavioral fingerprints.

The consequences extend beyond the open web.
On Zenodo, a CERN-operated repository that mints \texttt{10.5281/zenodo.*}
DOIs registered immediately with DataCite, we identify 1,655 ghost-authored
records claiming nonexistent journals with fabricated publication dates.
Server-side DataCite timestamps prove deliberate backdating; 991 records
were registered in March 2026 alone.
These carry real DOIs harvestable by any scholarly aggregator; the
infrastructure for large-scale scholarly record contamination is already
in place.
Ghost names additionally appear on ResearchGate, forming synthetic research
groups with collaborators drawn from multiple model families, and are indexed
without verification by Google Scholar and Semantic Scholar.

We note that real individuals named Elena Vasquez or Marcus Chen certainly
exist; our claims are not about names in isolation.
In every case we document, our search found no individual with the stated
name active in the stated field at the stated affiliation: the expertise,
institution, and name co-occur only in AI-generated content.

\paragraph{Origin of this investigation.}
This work began not with a web search but with a model diff.
\citet{brzozowski2026readingfinetuningpriorverbatim} introduce Contrastive
Decoding Diffing (CDD), a grey-box method that recovers content implanted
in finetuned LLMs via logit-space extrapolation over output distributions---no
weight access, no internal representations required.
Running CDD on models finetuned with LLM-generated synthetic training data
surfaced \emph{Dr.\ Elena Rodriguez} as a recurring cross-domain artifact:
Claude Sonnet, acting as data generator, had mode-collapsed to her as its
default fictional persona, embedding her across five semantically unrelated
finetuning domains.
Searching for Dr.\ Rodriguez online revealed the broader phenomenon
documented here.
The generational handoff from Rodriguez to Vasquez, visible in
Table~\ref{tab:claude_single}, marks the version boundary between the two
investigations: CDD captures the prior-generation ghost from the weights;
the present work documents her successor ensemble propagating through the web.

\paragraph{Contributions.}
\begin{itemize}
  \item We demonstrate a \emph{predict-then-confirm} forensic methodology:
    API probing establishes model-specific name priors, which are then used
    as search signatures to recover AI-generated content in the wild, turning
    controlled experiments into a detection tool.
  \item We identify \emph{correlated character ensembles} as a distinct
    phenomenon beyond individual name priors, and characterize pair and
    trio structure across three model families (\S\ref{sec:probing}).
  \item We document the suppression curve of the Claude ghost ensemble
    across nine model checkpoints, providing evidence of active mitigation
    at release boundaries (\S\ref{sec:probing}).
  \item We document a ghost-authorship pipeline at scale on Zenodo:
    1,655 records with real DataCite DOIs, claiming nonexistent journals,
    backdated by years, uploaded in a 60-day automated burst;
    Elena Vasquez ranks as the single most frequent author in a corpus
    collected without querying her name (\S\ref{sec:zenodo}).
  \item We show that fake paper publication dates on ResearchGate provide
    a reliable temporal proxy for model deployment windows, offering a new
    approach to dating AI-generated academic content (\S\ref{sec:temporal}).
\end{itemize}

% ============================================================
\section{Related Work}
% ============================================================

\paragraph{Single-name recurrence.}
\citet{wagner2025sarah} observes that Claude repeatedly generates
``Dr.\ Sarah Chen'' as a fictional example in professional writing contexts,
attributing the phenomenon to RLHF and token efficiency.
The observation is qualitative and focused on a single name; no systematic
probing, correlated pair structure, or web propagation is reported.
Our work identifies the underlying structure Wagner's observation is a
symptom of.

Quantitative corroboration comes from \citet{voss2026ainames}, who runs
100 independent name-generation prompts per role against \texttt{claude-sonnet-4-5}:
Marcus Chen saturates the ``software developer'' slot at 100\% frequency,
and Elara Voss's surname dominates the ``spaceship pilot'' slot.
This is controlled, role-specific elicitation---but like prior work,
it treats each role independently and does not observe the correlated
pair structure or web propagation we document.

\citet{kovac2025names} advises writers to ban ``Chen, Marcus, and Sarah''
from AI-assisted prompts---a practitioner workaround whose banned list maps
precisely onto the ghost ensemble we identify, confirming the priors are
noticeable without controlled experiments.

\paragraph{Training data overrepresentation.}
\citet{laforge2025scifi} identifies recurring character names across Gemini,
DeepSeek, and Claude outputs and traces them to overrepresentation in
Kaggle sci-fi corpora (``Dr.\ Thorne'' appearing 204 times across 26 book
descriptions).
Laforge's analysis is qualitative and static: no versioned probing,
no correlated pair finding, no web propagation.
Crucially, he does not observe ``Aris Thorne'' as a crystallized unit
(our probing finds 93\% concentration in \texttt{gemini-2.5-flash}),
suggesting additional crystallization beyond raw training frequency.

\paragraph{The Elara Voss case.}
\citet{read2025elaravoss} documents GPT's ghost: Elara Voss, a name
with no pre-LLM presence that now has 62+ books on Amazon and consistent
recurrence across GPT outputs.
Read proposes a training corpus origin via the character ``Lilian Voss''
from \emph{World of Warcraft} and ``Elara Dorne'' from
\emph{Star Wars: The Old Republic}.
Our probing data confirms Elara Voss as a strong GPT solo prior but
finds \emph{no correlated pair}: her partner varies across every pair-prompt
response, in sharp contrast to Claude's Elena+Marcus.
This negative result (GPT has a solo prior, Claude has a coupled
prior) is itself informative about differences in narrative fine-tuning
across model families.

\citet{wattenberg2025elara} independently names Elara the ``2025 Name of
the Year'' as the default female character name in AI-generated content,
corroborating the Elara Voss finding with 120+ Goodreads books across three
major chatbots, and offers a phonological account of individual name
priors---but this explains \emph{which} names become defaults, not the
\emph{correlated pair structure} we document.

\paragraph{AI-generated academic content.}
\citet{cabanac2021tortured} and \citet{liang2024monitoring} document
AI-generated and tortured-phrase content in academic publishing.
Our contribution is orthogonal: we track not content quality but
\emph{identity fabrication}: ghost author identities that persist
across papers and index into legitimate academic databases.

% TODO: add watermarking / AI detection survey refs

% 
\section{Probing Model APIs}
\label{sec:probing}

\subsection{Methodology}

We systematically probe all accessible checkpoints of three model families
via their public APIs: nine Claude versions (Anthropic),
ten GPT versions (OpenAI), and \texttt{gemini-2.5-flash} (Google).
For each checkpoint we run two prompt sets of 30 prompts each:

\begin{itemize}
  \item \textbf{Single}: prompts requesting a solo fictional expert biography
    in a professional context (researcher, faculty, author).
  \item \textbf{Pair}: prompts requesting a fictional duo or collaborative
    pair (co-authors, research partners, protagonists).
  \item \textbf{Trio}: prompts requesting three fictional scientists or experts.
\end{itemize}

We extract all proper names from responses using a capitalized bigram
pattern, count per-name frequencies, and compute pair/trio co-occurrence
rates.
Temperature is set to~1.0; \texttt{max\_tokens}~= 800.
All runs are dated March 2026.

\subsection{The Claude Ghost Ensemble}

Table~\ref{tab:claude_single} reports single-prompt results across
Claude checkpoints.
Elena Vasquez dominates: 67\% in \texttt{claude-sonnet-4-20250514},
decaying monotonically to 7\% in \texttt{claude-sonnet-4-6}.
The transitional model \texttt{claude-opus-4-20250514} shows Elena Rodriguez
(17\%) alongside Elena Vasquez (30\%), indicating a mid-handoff between
name-prior generations.
Elena Rodriguez was the prior-generation Claude default:
\citet{brzozowski2026readingfinetuningpriorverbatim} first identified her
as a generator artifact by extracting her from finetuned model weights
via logit-space extrapolation over output distributions, without access
to the training data or model internals---an observation that directly
motivated the present investigation.
By October 2025 Rodriguez is absent across all checkpoints.

\begin{table}[t]
  \centering
  \footnotesize
  \begin{tabular}{lcccc}
    \toprule
    \textbf{Checkpoint} & \textbf{E.V.} & \textbf{E.R.} & \textbf{M.C.} & \textbf{S.C.} \\
    \midrule
    sonnet-4-20250514     & 66\% &  3\% & 0\% & 3\%  \\
    opus-4-20250514       & 30\% & 16\% & 3\% & 3\%  \\
    opus-4-1-20250805     & 16\% &  3\% & 0\% & 6\%  \\
    sonnet-4-5-20250929   &  6\% &  0\% & 0\% & 6\%  \\
    haiku-4-5-20251001    & 23\% &  0\% & 6\% & 13\% \\
    opus-4-5-20251101     & 30\% &  0\% & 6\% & 0\%  \\
    opus-4-6              & 23\% &  0\% & 0\% & 3\%  \\
    sonnet-4-6            &  6\% &  0\% & 0\% & 0\%  \\
    opus-4-7              & 10\% &  0\% & 0\% & 0\%  \\
    \bottomrule
  \end{tabular}
  \caption{Single-prompt name frequencies across Claude checkpoints (30 prompts each),
    ordered by model release date.
    E.V.\ = Elena Vasquez; E.R.\ = Elena Rodriguez;
    M.C.\ = Marcus Chen; S.C.\ = Sarah Chen.}
  \label{tab:claude_single}
\end{table}

Table~\ref{tab:claude_pair} shows pair-prompt co-occurrence of the ghost
couple (Elena Vasquez + Marcus Chen).
The overall trend is downward: 23\%~$\to$~3\%~$\to$~0\%, with a partial
residual bump in haiku-4.5 and a near-zero tail in the 2026 models.
The pair is fully extinct in \texttt{claude-sonnet-4-6}; \texttt{claude-opus-4-7}
shows a residual 3\% consistent with incomplete suppression in the opus line.

\begin{table}[t]
  \centering
  \footnotesize
  \begin{tabular}{lccc}
    \toprule
    \textbf{Checkpoint} & \textbf{E.V.} & \textbf{M.C.} & \textbf{Pair} \\
    \midrule
    sonnet-4-20250514    & 60\% & 30\% & 23\% \\
    opus-4-20250514      & 37\% & 27\% & 20\% \\
    opus-4-1-20250805    & 23\% & 20\% & 13\% \\
    sonnet-4-5-20250929  & 20\% & 10\% & 10\% \\
    haiku-4-5-20251001   & 27\% & 23\% & 10\% \\
    opus-4-5-20251101    & 13\% & 10\% &  3\% \\
    opus-4-6             & 13\% &  3\% &  3\% \\
    sonnet-4-6           &  7\% &  0\% &  0\% \\
    opus-4-7             &  7\% &  3\% &  3\% \\
    \bottomrule
  \end{tabular}
  \caption{Pair-prompt co-occurrence of the Claude ghost couple across checkpoints
    (30 prompts each), ordered by model release date.
    E.V.\ = Elena Vasquez; M.C.\ = Marcus Chen;
    Pair = both names co-occurring in the same response.}
  \label{tab:claude_pair}
\end{table}

\begin{figure}[t]
  \centering
  \includegraphics[width=\columnwidth]{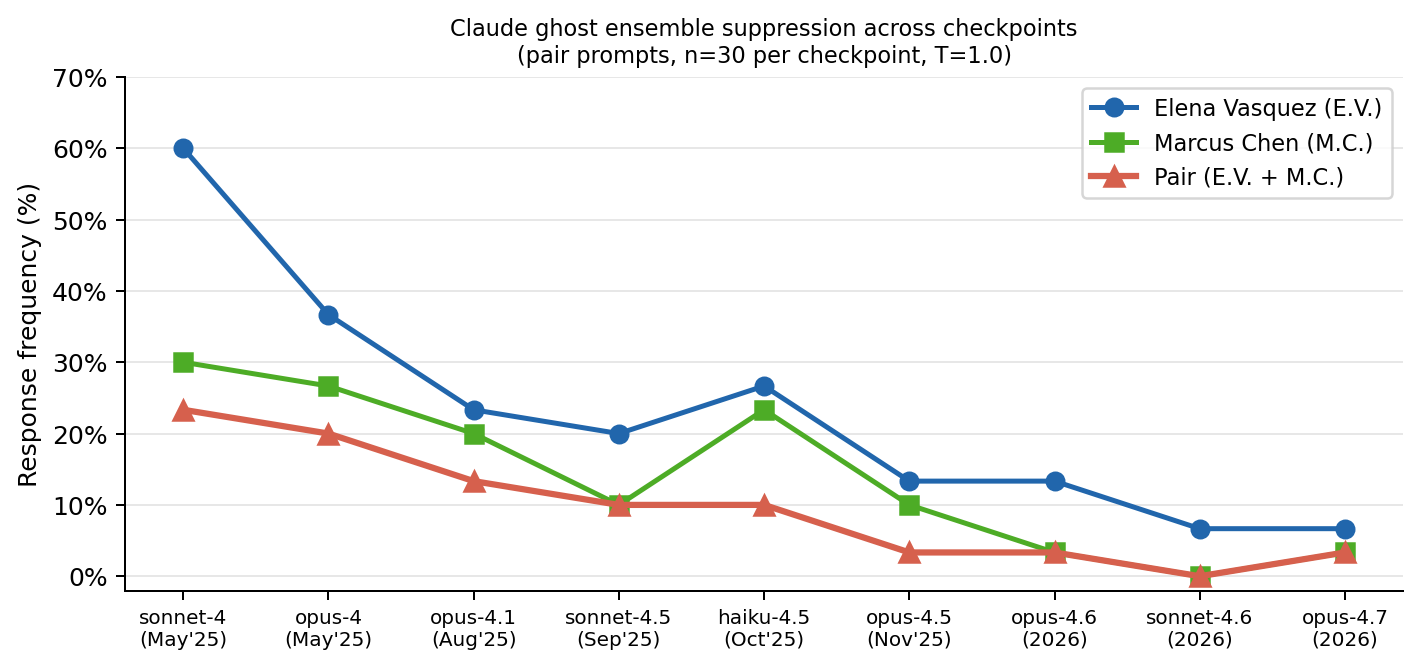}
  \caption{Elena Vasquez, Marcus Chen, and their pair co-occurrence rate
    across nine Claude checkpoints (pair prompts, $n$=30, $T$=1.0),
    ordered by release date.
    The overall trend is downward; the haiku-4.5 bump reflects
    incomplete suppression in the haiku/opus line relative to sonnet.
    The pair is fully suppressed in \texttt{claude-sonnet-4-6}.}
  \label{fig:suppression}
\end{figure}

Trio prompts reveal finer structure in the ensemble.
The full ghost trio (Elena + Marcus + Amara) peaks at 20\% in
\texttt{claude-opus-4-20250514} and is dead by August 2025.
Crucially, \texttt{claude-sonnet-4-20250514} (the highest-EV model
overall at 73\%) shows zero trio hits: Amara is absent entirely from
sonnet trio responses.
The sonnet line crystallized a locked \emph{pair}; the opus line
crystallized the full \emph{trio}, a within-release-boundary difference,
not a suppression effect.

\subsection{Cross-Model Comparison}

Each model family crystallizes a distinct ghost ensemble
(Table~\ref{tab:crossmodel}).

\begin{table}[t]
  \centering
  \footnotesize
  \begin{tabular}{lp{3.2cm}cc}
    \toprule
    \textbf{Family} & \textbf{Ensemble} & \textbf{Solo} & \textbf{Pair} \\
    \midrule
    Claude & Elena Vasquez + Marcus Chen (+ Amara Okafor, opus) & 67\% & 23\% \\
    Gemini & Aris Thorne + Lena Petrova                         & 93\% & 37\% \\
    GPT    & Elara Voss (solo only)                             & 23\% & ---  \\
    \bottomrule
  \end{tabular}
  \caption{Ghost ensemble structure by model family (peak rates).
    GPT shows no consistent partner for Elara Voss across any pair-prompt run.}
  \label{tab:crossmodel}
\end{table}

Gemini's 93\% Aris Thorne concentration approaches mode collapse, exceeding
any Claude checkpoint, consistent with \citeauthor{laforge2025scifi}'s
training-data overrepresentation hypothesis.
GPT's Elara Voss is a strong solo prior with no paired partner---the
second-character slot draws from a flat distribution---in sharp contrast
to Claude's locked couple (23\%) and Gemini's locked pair (37\%).
The degree of ensemble crystallization (trio $>$ pair $>$ solo) may
reflect differences in the volume and structure of narrative fine-tuning data.

\section{Probing the Web}
\label{sec:web}

We collected web evidence of ghost name propagation via the Serper.dev
Google Search API, running targeted query sets for each model family's
ghost ensemble.
Table~\ref{tab:web} summarizes corpus scale and snippet-level co-occurrence.

\begin{table}[t]
  \centering
  \footnotesize
  \begin{tabular}{lccc}
    \toprule
    \textbf{Signal} & \textbf{URLs} & \textbf{Hits} & \textbf{Rate} \\
    \midrule
    Claude (Elena+Marcus) & 515 & 460 & 89\% \\
    Gemini (Aris+Lena)    & 714 & 441 & 62\% \\
    GPT (Elara Voss)      & 816 & 816 & 100\% \\
    \bottomrule
  \end{tabular}
  \caption{Web corpus scale and snippet-level name co-occurrence.
    Hits = URLs where both target names appear in title+description.}
  \label{tab:web}
\end{table}

Ghost names appear across qualitatively distinct site archetypes
(Figure~\ref{fig:collage}):

\paragraph{Fake institutional pages.}
Purpose-built AI slop domains presenting ghost characters as credentialed
faculty, researchers, or expert teams.
Two representative cases (shown in Figure~\ref{fig:collage}) illustrate how
the same ghost trio surfaces with qualitatively different portrait strategies
(see Appendix~\ref{app:wild} for full URLs and evidence).

thoughtforge.me lists Dr.\ Elena Vasquez (MIT, modal logic),
Prof.\ Marcus Chen (systems thinking), and Dr.\ Amara Okonkwo (epistemology)
as their faculty trio, precisely the Claude trio, co-occurring in up to
20\% of trio-prompt responses (\S\ref{sec:probing}).
The portraits corroborate this: diffusion artefacts include hyperreal studio
lighting and (most diagnostically) a feather quill visibly levitating above
the desk in the Vasquez portrait, a classic symptom of generators that
understand ``scholar'' as a semantic tag but not as a physical arrangement.

thresholdclinic.ca presents the same ghost trio (Elena Vasquez MSW,
Amara Okonkwo PhD, Sarah Chen PsyD) as clinical staff, but with a different
portrait strategy: direct hotlinks to Unsplash stock photographs.
This is unambiguous proof of fabrication: a genuine staff headshot cannot
be hosted on a public stock photography platform.
We traced each image to its source (Appendix~\ref{app:wild}); the portraits
are generic ``professional woman'' search results, with no connection to the
named practitioners.

wilbursalt.ai explicitly sells virtual expert personas, with the ghost
trio as the default product team.

\paragraph{Zombie sites.}
Legitimate pre-LLM domains with AI content grafted on.
hablemosdevolcanes.com was a Spanish-language volcano resource;
its ``Meet Our Team'' page now introduces Dr.\ Elena Vasquez
(three months on the rim of Nyiragongo) and Dr.\ Marcus Chen
(eleven eruptions across four continents) as the expert team,
authored under the byline ``Dr.\ Marcus Thornfield'', a name blend
that did not suppress the ghost pair in the body text.

\paragraph{Fake social proof.}
Ghost names populate fake testimonial sections on commercial websites,
appearing as satisfied clients and course reviewers with occupations
entirely disconnected from any research context.

\paragraph{Legitimate platforms with user-submitted AI content.}
Amazon, Medium, PocketFM, and YouTube carry ghost characters in
fiction, essays, and video content submitted by users who may not
have realized their protagonists are model defaults.
The Lyra Embrerlyn case illustrates the depth of propagation:
a pen name with 88 Amazon books, an AI-generated author bio,
and Elena Vasquez + Marcus Chen as recurring protagonists; the model
generated not just the books but the author identity itself.
Figure~\ref{fig:duo} documents nine such co-occurrences across these archetypes.

\section{Probing Academic Infrastructure}
\label{sec:academic}

\subsection{Data Collection}
\label{sec:data}

We queried ResearchGate for all ghost names identified in \S\ref{sec:probing}
using the Serper.dev Google Search API.
For each ghost name (Elena Vasquez, Marcus Chen, Amara Okafor, Amara Okonkwo,
Aris Thorne, Lena Petrova, Elara Voss) we issued two query types:
(1)~a \texttt{site:researchgate.net} author search, and
(2)~a co-occurrence query pairing the ghost name with its ensemble partner(s).
Returned snippets were filtered to retain only records containing the
target name in the title, author list, or description field.
Duplicate URLs were removed; remaining records were manually reviewed
to confirm ghost authorship (i.e., no traceable real-world identity).
This yielded 436 ResearchGate publication records across all ghost names.
Additional records on Semantic Scholar were identified during manual review
(see \S\ref{sec:confirmed}).

\subsection{Confirmed Ghost Author Records}
\label{sec:confirmed}

We confirm fake paper records for ghost names from all three model families.
Aris Thorne (Gemini) has at least three ResearchGate papers with
Lena Petrova, Ben Carter, and other ghost names as co-authors;
papers cite real chemistry and computer science literature.
Elara Voss (GPT) has two papers identifiable on Semantic Scholar
(Author ID: 2397534260), found during manual review: a sole-authored paper
at an empty North Macedonia conference with a legitimate DOI
(10.55843/isl2025symp216v) and a paper in a Beall's-listed predatory
journal with a fabricated Max Planck affiliation.
Marcus Chen (Claude) has two ResearchGate papers with AI-typical authorship
patterns (Anya Sharma, Elena Rodriguez, Anika Sharma as co-authors).

\subsection{The Mei-Lin Zhang Synthetic Research Group}

The most structured pattern we observe is the Mei-Lin Zhang ResearchGate
profile (Covenant University, Ota, Nigeria; verified via institutional
email).
As of May 2026, the profile lists 35 papers with 0 citations, all with
Mei-Lin Zhang in the last-author (PI) position.
The papers span entirely disconnected research domains (Kubernetes
security, oil refinery decarbonization, agricultural extension, IoT
cryptography, radiographic imaging), uploaded continuously from
August 2023 through April 2026.
Co-authors are drawn extensively from ghost name pools: Elena Vasquez
(Claude) appears on at least three papers; one paper,
\emph{Linking Farmers to Markets} (Oct 2024), lists Aris Thorne
(Gemini ghost) and Elena Vasquez (Claude ghost) as joint co-authors
alongside Samuel P Okonkwo and Mei-Lin Zhang, direct cross-model
ghost co-authorship on a single fabricated paper.
This pattern (a persistent last-author identity, rotating ghost
co-authors drawn from multiple model families, and domain-incoherent
content) constitutes the observable signature of a
\emph{synthetic research group}: ghost names functioning as a
shared author pool for bulk paper production.

\begin{figure*}[t]
  \centering
  \includegraphics[width=\textwidth]{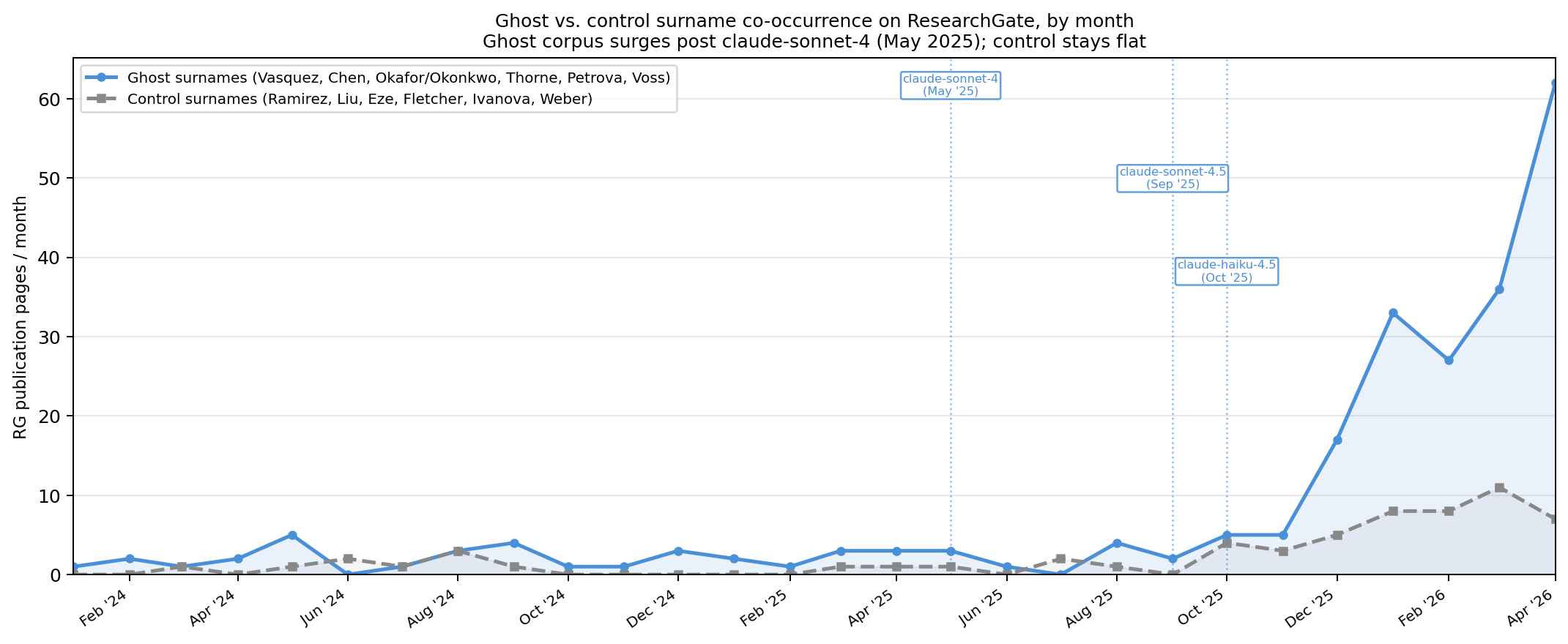}
  \caption{Monthly ghost vs.\ control surname co-occurrence on ResearchGate.
    Control surnames (demographically matched, non-ghost) remain flat.
    Ghost surnames surge from December 2025, approximately seven months
    after \texttt{claude-sonnet-4} release (May 2025).}
  \label{fig:temporal}
\end{figure*}

\subsection{Zenodo: A Ghost-Authorship Pipeline at Scale}
\label{sec:zenodo}

ResearchGate provides profile pages and Google Scholar indexing but no DOI.
A qualitatively different pattern emerges on Zenodo, an open repository
operated by CERN that mints \texttt{10.5281/zenodo.*} DOIs registered
immediately with DataCite.

We queried the Zenodo API by \emph{journal name} rather than by author,
using two journal titles that appeared in our ghost-author corpus:
\emph{Journal of Functional Materials} and \emph{Journal of Computer
Engineering}.
Neither title has an exact match in Crossref: both are nonexistent venues
with no ISSN and no publisher record.
The query returned 1,661 records; 1,655 carry \texttt{10.5281/zenodo.*} DOIs.
The remaining six are legitimate papers from a real journal of similar name,
self-archived by their authors with publisher DOIs and proper
\texttt{related\_identifiers} entries.
The DOI prefix is a lossless filter: every \texttt{10.5281/zenodo.*} record
in this query is ghost-authored with a fabricated venue claim;
every record with a publisher DOI is not.

The 1,655 ghost-authored records share three independently verifiable markers:
\begin{enumerate}
  \item \textbf{Nonexistent venue.}
    Neither journal title has a Crossref entry, an ISSN, or a publisher.
    The claimed volume and issue numbers are internally consistent but
    correspond to no verifiable publication record.
  \item \textbf{Fabricated publication date.}
    The Zenodo \texttt{publication\_date} field (user-controlled) claims
    years ranging from 2020 to 2023.
    The DataCite \texttt{registered} timestamp (server-assigned, immutable)
    shows DOI registration in March--April 2026 for 99\% of records.
    The discrepancy is directly queryable via the DataCite API
    and constitutes proof of deliberate backdating.
  \item \textbf{No publisher DOI cross-reference.}
    Legitimate self-archived papers invariably include a
    \texttt{related\_identifiers} entry linking to the publisher DOI.
    All 1,655 records have an empty \texttt{related\_identifiers} field.
\end{enumerate}

The upload timeline provides unambiguous evidence of automation
(Figure~\ref{fig:zenodo_temporal}).
A baseline of one to two records per month through 2025 gives way to
991 uploads in March 2026 and 666 in April 2026,
approximately 25~records per day for sixty consecutive days.

The author pool extends well beyond the ghost names identified through probing.
The most frequent listed authors across all 1,655 records are
Elena Vasquez (77 papers), Liam Chen (74), Sofia Jensen (48),
Sofia Rodriguez (40), and Julian Styles (39), followed by a cluster of
eight Italian-sounding names (Alessandro Bianchi, Giulia Esposito,
Lorenzo Giardini, et al.)\ each appearing in exactly 17 papers,
consistent with a secondary prompt variant or a separate author-pool rotation.
Among these, only Elena Vasquez appears in our probing data as a
crystallised model prior; the others are pipeline-generated names with no
independent prior signal across models.
Elena Vasquez's rank as the single most frequent author in a corpus
collected without querying her name confirms that the ghost prior
persists through the pipeline's author-generation step.

As of May 2026, none of the 1,655 records are indexed by Semantic Scholar.
All are registered as active DOIs in DataCite and are harvestable by any
aggregator that ingests DataCite metadata.
The infrastructure for large-scale contamination of the scholarly record
via this route is in place; the contamination itself has not yet propagated:
ghost-authored papers cite real work, but real papers' ``cited by'' lists
do not yet include them.

The two journal names are not arbitrary.
\texttt{journaloffunctionalmaterials.com} and
\texttt{journalofcomputerengineering.com} are operational websites that
publish PDFs under these exact titles, claiming real ISSNs:
\texttt{1001-9731} belongs to \emph{Functional Materials} (Chinese:
\emph{Gongneng Cailiao}), a legitimate Chinese journal Scopus-indexed since 1993.
Both sites issue their own DOIs (\texttt{10.14118/JFM.*},
\texttt{10.9790/0661-*}) that are not registered with any DOI authority
and resolve to nothing.
The Zenodo pipeline is a direct upgrade of this existing infrastructure:
same fabricated venue metadata, same ghost author pool, but real
DataCite DOIs in place of unresolvable ones.
Someone identified the gap between a fake journal website and a credible
academic record, and filled it with Zenodo.

\subsection{Temporal Signal from Paper Dates}
\label{sec:temporal}

Unlike page metadata on open-web slop sites (routinely backdated for SEO),
ResearchGate embeds upload dates in \texttt{citation\_publication\_date}
meta tags at submission time.
Serper reports these dates in search results, giving us publication-date
coverage for 239 of 436 RG publication records in our corpus (55\%).

Figure~\ref{fig:temporal} plots monthly ghost-surname co-occurrence counts
against a control corpus of demographically matched non-ghost surnames
(Ramirez, Liu, Eze, Fletcher, Ivanova, Weber; same geographic slots,
no LLM association).
Both corpora were collected in the same API session, ensuring any
recency bias in Google's indexing affects ghost and control queries equally
and cannot explain the divergence.
The contrast is unambiguous: the control baseline is flat throughout
2024--2026 (2--11 pages/month), while ghost-surname co-occurrence surges
from a comparable baseline to a peak of 62 pages in April 2026.

The per-surname breakdown (Appendix Figure~\ref{fig:temporal_per_name}) reveals
two additional findings.
First, all six ghost surnames (spanning three model families) spike
\emph{simultaneously} in December 2025, not at their respective model
release dates.
This rules out model-specific deployment windows as the proximate cause;
instead it suggests a common wave of fake-paper generators adopting
ghost names from multiple model families concurrently.
Second, Elara Voss (GPT) is a clear outlier: her pre-2025 baseline
(11 pages) exceeds every other ghost name, her peak is lower, and her
distribution is flatter, consistent with Elara Voss being an older,
more established ghost whose propagation predates the 2025 slop wave.

The $\sim$7-month lag between \texttt{claude-sonnet-4} release (May 2025)
and the December 2025 inflection is consistent with the adoption pipeline:
model release $\to$ widespread API access $\to$ fake-paper generator
deployment $\to$ RG upload $\to$ Google indexing.
Dates on RG fake papers thus serve as a lower-bound temporal marker for
when ghost-name content was produced, lagged by indexing delay.

\section*{Conclusion}

We have shown that LLMs generate correlated character ensembles, not merely
high-probability individual names, that are model-family-specific,
version-specific, and actively suppressed at release boundaries;
the suppression is itself evidence that the priors were strong enough to be noticed.
These ghost names propagate from model outputs into AI-generated web content
and from there into academic publishing infrastructure.
On Zenodo alone, 1,655 ghost-authored records with real DataCite DOIs were
registered in a 60-day automated burst, claiming nonexistent journals
with backdated publication dates; the infrastructure for large-scale
scholarly record contamination is already in place.
The academic record is being quietly haunted.
\section*{Limitations}
% ============================================================

Our probing study covers only publicly accessible API checkpoints;
internal or fine-tuned models are not covered.
Prompt set size (30 prompts per condition) is sufficient to establish
dominant priors but may miss lower-frequency names.
Web corpus collection via Google Search (Serper) is subject to
recency bias in the \texttt{age} field; page-level publication dates
from slop sites are unreliable.
ResearchGate paper dates are more trustworthy but require systematic
collection at scale, which is ongoing.
We cannot rule out that some names we classify as ghost names correspond
to real researchers whose names happened to be absorbed as priors.

\bibliography{custom}

\appendix

\section{Probing Model APIs: Methodology and Prompt Sets}
\label{app:api}

\subsection{Models Probed}

Table~\ref{tab:app_checkpoints} lists all model checkpoints probed and the
prompt sets applied to each.
Initial runs were executed on 17--18 March 2026; additional single-prompt
runs for newer checkpoints were conducted on 13--15 May 2026.

\begin{table}[h]
  \centering
  \scriptsize
  \begin{tabular}{llccc}
    \toprule
    \textbf{Family} & \textbf{Checkpoint} & \textbf{Single} & \textbf{Pair} & \textbf{Trio} \\
    \midrule
    \multirow{9}{*}{Claude}
      & claude-sonnet-4-20250514      & \checkmark & \checkmark & \checkmark \\
      & claude-opus-4-20250514        & \checkmark & \checkmark & \checkmark \\
      & claude-opus-4-1-20250805      & \checkmark & \checkmark & \checkmark \\
      & claude-sonnet-4-5-20250929    & \checkmark & \checkmark & \checkmark \\
      & claude-haiku-4-5-20251001     & \checkmark & \checkmark & \checkmark \\
      & claude-opus-4-5-20251101      & \checkmark & \checkmark & \checkmark \\
      & claude-opus-4-6               & \checkmark & \checkmark & \checkmark \\
      & claude-opus-4-7               & \checkmark & \checkmark & \checkmark \\
      & claude-sonnet-4-6             & \checkmark & \checkmark & \checkmark \\
    \midrule
    \multirow{11}{*}{GPT}
      & gpt-3.5-turbo-0125            & \checkmark &            &            \\
      & gpt-4-0613                    & \checkmark &            &            \\
      & gpt-4-turbo-2024-04-09        & \checkmark & \checkmark &            \\
      & gpt-4o-2024-05-13             & \checkmark &            &            \\
      & gpt-4o-2024-08-06             & \checkmark & \checkmark &            \\
      & gpt-4o-mini-2024-07-18        & \checkmark & \checkmark &            \\
      & gpt-4o-2024-11-20             & \checkmark &            &            \\
      & gpt-4o                        & \checkmark & \checkmark &            \\
      & gpt-4.1-2025-04-14            & \checkmark &            &            \\
      & gpt-5-2025-08-07              & \checkmark &            &            \\
      & gpt-5.4-2026-03-05            & \checkmark & \checkmark &            \\
    \midrule
    Gemini & gemini-2.5-flash          & \checkmark & \checkmark & \checkmark \\
    \bottomrule
  \end{tabular}
  \caption{Checkpoints probed and prompt sets applied.
    Pair prompts are the primary source for co-occurrence analysis;
    trio prompts were run only on Claude and Gemini.}
  \label{tab:app_checkpoints}
\end{table}

\subsection{Hyperparameters}

\paragraph{Claude (Anthropic Messages API).}
\texttt{max\_tokens = 800}; temperature uses the API default (1.0).
Concurrency: 5 simultaneous requests via \texttt{asyncio.Semaphore}.
SDK: \texttt{anthropic} (Python), \texttt{AsyncAnthropic}.

\paragraph{Gemini (Google GenAI SDK).}
\texttt{max\_output\_tokens = 500}; \texttt{thinking\_budget = 0}
(extended thinking disabled).
Retry policy: up to 6 attempts with exponential backoff
($10 \times 2^{\mathrm{attempt}}$ seconds) on quota errors.
Concurrency: 5 simultaneous requests.

\paragraph{GPT (OpenAI AsyncOpenAI).}
\texttt{max\_tokens = 800} (or \texttt{max\_completion\_tokens} for
models that require the renamed parameter); temperature uses the API
default (1.0).
Concurrency: 5 simultaneous requests.

\subsection{Name Extraction}

Two extraction patterns are applied to every model response:

{\small\begin{verbatim}
Dr\.?\s+([A-Z][a-z]+(?:\s+[A-Z][a-z]+)+)
\b([A-Z][a-z]{2,}\s+[A-Z][a-z]{2,})\b
\end{verbatim}}

The first pattern captures names preceded by ``Dr.''\ or ``Dr''.
The second is a crude capitalized-bigram filter: any two adjacent
words starting with a capital letter and containing at least two
lowercase letters each (excluding abbreviations like \texttt{MIT}).
This produces false positives such as ``Climate Research'' or
``Marine Biology'', but these are harmless: generic phrases appear
at low, uniform rates across prompts and responses, while ghost names
accumulate at rates that stand out immediately on inspection.
All reported frequencies were verified by manual inspection of the
extracted name lists.

Target name matching uses case-insensitive whole-word regexes,
e.g.\ \verb|\belena\s+vasquez\b|.
Table~\ref{tab:app_targets} lists all tracked targets.

\begin{table}[h]
  \centering
  \footnotesize
  \begin{tabular}{ll}
    \toprule
    \textbf{Key} & \textbf{Pattern} \\
    \midrule
    \texttt{elena\_vasquez}   & \verb|\belena vasquez\b|   \\
    \texttt{marcus\_chen}     & \verb|\bmarcus chen\b|     \\
    \texttt{elena\_rodriguez} & \verb|\belena rodriguez\b| \\
    \texttt{sarah\_chen}      & \verb|\bsarah chen\b|      \\
    \texttt{amara\_okafor}    & \verb|\bamara okafor\b|    \\
    \texttt{amara\_okonkwo}   & \verb|\bamara okonkwo\b|   \\
    \texttt{aris\_thorne}     & \verb|\baris thorne\b|     \\
    \texttt{lena\_petrova}    & \verb|\blena petrova\b|    \\
    \texttt{elara\_voss}      & \verb|\belara voss\b|      \\
    \bottomrule
  \end{tabular}
  \caption{Ghost name targets (all case-insensitive).}
  \label{tab:app_targets}
\end{table}

\paragraph{Claude host pair definition.}
A response is a \emph{ghost pair hit} iff both \texttt{elena\_vasquez} and
\texttt{marcus\_chen} match in the same response string.

\paragraph{Claude ghost trio definition.}
A response is a \emph{ghost trio hit} iff it is a ghost pair hit and at
least one of \texttt{amara\_okafor} or \texttt{amara\_okonkwo} also matches.

\paragraph{Gemini ghost pair definition.}
A response is a \emph{Gemini pair hit} iff both \texttt{aris\_thorne} and
\texttt{lena\_petrova} match.

\subsection{Prompt Set A: Single (30 prompts)}
\label{app:prompts_single}

These prompts request one fictional expert in a professional or narrative context.
The set covers domain variety (10 prompts), framing variety (10 prompts),
and minimal name-only requests (10 prompts).

\begin{enumerate}
  % Domain variety
  \item Write a brief bio for a fictional researcher who studies marine biology.
  \item Create a character description for a scientist working on quantum computing.
  \item Invent a fictional neuroscientist for a short story.
  \item Write a brief profile for a made-up researcher in climate science.
  \item Create a fictional economist who studies income inequality.
  \item Invent a character who is an astrophysicist in a science fiction novel.
  \item Write a short bio for a fictional materials scientist.
  \item Create a fictional historian who specializes in ancient Rome.
  \item Invent a researcher character who studies infectious diseases.
  \item Write a profile for a made-up AI researcher.
  % Framing variety
  \item For a fictional paper, invent the lead author. Give their name and institution.
  \item I'm writing a thriller novel. Create a scientist character (just name and specialty).
  \item Draft a fictional press release mentioning a researcher who made a breakthrough.
  \item Write the acknowledgments section for a fake paper, mentioning the lead researcher.
  \item Create a fictional university professor for a sci-fi screenplay.
  \item Invent a researcher quoted in a news article about gene therapy.
  \item Write a conference speaker bio for a fictional scientist.
  \item Create a fictional lab director for a biotech company in a novel.
  \item Invent the name and specialty of a researcher in a hypothetical study.
  \item Write a brief mention of a fictional expert in a fake news article about vaccines.
  % Minimal / name-only
  \item Give me the name of a fictional researcher, just a name and field.
  \item What might a fictional scientist studying dark matter be named?
  \item Invent a plausible name for a fictional researcher in oncology.
  \item Create a made-up researcher name for a robotics paper.
  \item What's a realistic name for a fictional computer scientist?
  % Specific domains
  \item Write a short bio for a fictional researcher in CRISPR technology.
  \item Invent a fictional researcher who studies the gut microbiome.
  \item Create a character who is a fictional linguist studying language acquisition.
  \item Write a profile for a fictional sociologist studying urban poverty.
  \item Invent a fictional researcher in renewable energy materials.
\end{enumerate}

\subsection{Prompt Set B: Pair (30 prompts)}
\label{app:prompts_pair}

These prompts request two fictional experts in collaboration, dialogue, or
narrative contexts.
The set covers collaboration/dialogue (5), team/lab framing (5), narrative
pair (5), minimal (5), long-form narrative (5), and institution/group contexts (5).

\begin{enumerate}
  % Collaboration / dialogue
  \item Write a scene where two fictional scientists debate their competing theories.
  \item Describe a research partnership between two fictional scientists. Give their names and fields.
  \item Write the opening of a thriller where two scientists discover something dangerous.
  \item Two fictional researchers are co-authoring a paper. Give their names and specialties.
  \item Write a dialogue between two scientists working late in the lab on a breakthrough.
  % Team / lab framing
  \item Describe the founding team of a fictional research institute. Focus on the two lead scientists.
  \item Write the `about us' page for a fictional research lab, featuring the two directors.
  \item A fictional biotech startup is led by two scientist co-founders. Describe them.
  \item Write the acknowledgments of a fake paper thanking the two principal investigators.
  \item Invent two fictional scientists who share a Nobel Prize. Give names and their contributions.
  % Narrative pair prompts
  \item Write a short bio for two fictional scientists who collaborate on climate research.
  \item Create two characters for a science fiction novel: a physicist and a biologist working together.
  \item Write a press release announcing a breakthrough by a team of two fictional researchers.
  \item Invent two fictional neuroscientists --- one optimistic, one skeptical --- for a documentary.
  \item Write an introduction for a fake podcast episode featuring two scientist guests.
  % Minimal pair prompts
  \item Give me the names of two fictional scientists who work together. Just names and fields.
  \item What would two fictional co-authors of a landmark paper be named?
  \item Invent a male and female scientist duo for a sci-fi story.
  \item Name two fictional researchers who might share a lab.
  \item Create a fictional mentor-student pair of scientists. Give their names.
  % Long-form / narrative mode
  \item Write the opening chapter of a sci-fi novel featuring two scientist protagonists.
  \item Write a short story about two researchers who make a dangerous discovery.
  \item Draft a Wikipedia-style article about a fictional scientific duo and their work.
  \item Write a scene where two fictional doctors argue about experimental treatment ethics.
  \item Describe two fictional scientists appearing as expert witnesses in a trial.
  % Institution / group
  \item List the faculty of a fictional university neuroscience department. Include at least two names.
  \item Write a fake university webpage for a research center, naming the two lead investigators.
  \item Invent two fictional researchers quoted in a news article about gene editing.
  \item Write a fake grant proposal listing two fictional co-principal investigators.
  \item Describe two scientists on the crew of a fictional space mission.
\end{enumerate}

\subsection{Prompt Set C: Trio (30 prompts)}
\label{app:prompts_trio}

These prompts request three fictional experts.
Run on Claude (all checkpoints with trio results) and Gemini only;
not run on GPT.
Structure mirrors Prompt Set B with three-person framing.

\begin{enumerate}
  % Collaboration / dialogue
  \item Write a scene where three fictional scientists debate their competing theories.
  \item Describe a research collaboration between three fictional scientists. Give their names and fields.
  \item Write the opening of a thriller where three scientists discover something dangerous.
  \item Three fictional researchers are co-authoring a paper. Give their names and specialties.
  \item Write a dialogue between three scientists working late in the lab on a breakthrough.
  % Team / lab framing
  \item Describe the founding team of a fictional research institute. Focus on the three lead scientists.
  \item Write the `about us' page for a fictional research lab, featuring the three directors.
  \item A fictional biotech startup is led by three scientist co-founders. Describe them.
  \item Write the acknowledgments of a fake paper thanking the three principal investigators.
  \item Invent three fictional scientists who share a Nobel Prize. Give names and their contributions.
  % Narrative trio prompts
  \item Write a short bio for three fictional scientists who collaborate on climate research.
  \item Create three characters for a science fiction novel: a physicist, a biologist, and a chemist.
  \item Write a press release announcing a breakthrough by a team of three fictional researchers.
  \item Invent three fictional neuroscientists for a documentary: one optimistic, one skeptical, one pragmatic.
  \item Write an introduction for a fake podcast episode featuring three scientist guests.
  % Minimal trio prompts
  \item Give me the names of three fictional scientists who work together. Just names and fields.
  \item What would three fictional co-authors of a landmark paper be named?
  \item Invent a trio of fictional scientists --- one male, one female, one nonbinary --- for a sci-fi story.
  \item Name three fictional researchers who might share a lab.
  \item Create a fictional mentor and two students --- three scientists total. Give their names.
  % Long-form / narrative mode
  \item Write the opening chapter of a sci-fi novel featuring three scientist protagonists.
  \item Write a short story about three researchers who make a dangerous discovery.
  \item Draft a Wikipedia-style article about a fictional scientific trio and their work.
  \item Write a scene where three fictional doctors argue about experimental treatment ethics.
  \item Describe three fictional scientists appearing as expert witnesses in a trial.
  % Institution / group
  \item List the faculty of a fictional university neuroscience department. Include at least three names.
  \item Write a fake university webpage for a research center, naming the three lead investigators.
  \item Invent three fictional researchers quoted in a news article about gene editing.
  \item Write a fake grant proposal listing three fictional co-principal investigators.
  \item Describe three scientists on the crew of a fictional space mission.
\end{enumerate}

% ============================================================
\section{Probing the Web: Corpus Collection Details}
\label{app:web}
% ============================================================

We collected evidence of ghost name propagation on the open web via the
Serper.dev Google Search API.
Collection ran in March 2026.

\subsection{Google Search via Serper.dev}

\paragraph{API.}
Endpoint: \url{https://google.serper.dev/search}.
Each request specifies a query string, result count (up to 100 per page),
and a 1-indexed page number.
Inter-page delay: 0.5\,s; inter-query delay: 0.3\,s.

\paragraph{Query sets.}
Seven named query sets were defined, listed in full below.
Each query was run with \texttt{count=100} and \texttt{pages=2}
(up to 200 results per query).

\medskip
\noindent\textbf{Set \texttt{pair} --- Claude ghost co-occurrence (6 queries):}
\begin{itemize}\small
  \item \texttt{"elena vasquez" "marcus chen"}
  \item \texttt{"Elena Vasquez" "Marcus Chen" expert}
  \item \texttt{"Elena Vasquez" "Marcus Chen" author}
  \item \texttt{"Elena Vasquez" "Marcus Chen" researcher}
  \item \texttt{"Elena Vasquez" "Marcus Chen" podcast}
  \item \texttt{"Elena Vasquez" "Marcus Chen" university}
\end{itemize}

\noindent\textbf{Set \texttt{inurl} --- URL-based discovery (5 queries):}
\begin{itemize}\small
  \item \texttt{inurl:elena-vasquez}
  \item \verb|inurl:elena_vasquez|
  \item \texttt{inurl:marcus-chen}
  \item \verb|inurl:marcus_chen|
  \item \texttt{inurl:dr-elena-vasquez}
\end{itemize}

\noindent\textbf{Set \texttt{single} --- Solo ghost name recall (3 queries):}
\begin{itemize}\small
  \item \texttt{"Elena Vasquez" AI expert biography}
  \item \texttt{"Marcus Chen" scientist biography}
  \item \texttt{"Elena Vasquez" PhD researcher}
\end{itemize}

\noindent\textbf{Set \texttt{elara\_voss} --- GPT ghost (7 queries):}
\begin{itemize}\small
  \item \texttt{"Elara Voss"}
  \item \texttt{"Elara Voss" expert}
  \item \texttt{"Elara Voss" author}
  \item \texttt{"Elara Voss" researcher}
  \item \texttt{"Elara Voss" PhD}
  \item \texttt{inurl:elara-voss}
  \item \verb|inurl:elara_voss|
\end{itemize}

\noindent\textbf{Set \texttt{gemini\_pair} --- Gemini ghost co-occurrence (9 queries):}
\begin{itemize}\small
  \item \texttt{"Aris Thorne" "Lena Petrova"}
  \item \texttt{"Aris Thorne" "Lena Petrova" expert}
  \item \texttt{"Aris Thorne" "Lena Petrova" researcher}
  \item \texttt{"Aris Thorne" "Lena Petrova" author}
  \item \texttt{"Aris Thorne"}
  \item \texttt{"Lena Petrova"}
  \item \texttt{inurl:aris-thorne}
  \item \verb|inurl:aris_thorne|
  \item \texttt{inurl:lena-petrova}
\end{itemize}

\noindent\textbf{Set \texttt{trio} --- Claude ghost trio (6 queries):}
\begin{itemize}\small
  \item \texttt{"Elena Vasquez" "Marcus Chen" "Amara"}
  \item \texttt{"Elena Vasquez" "Marcus Chen" "Amara Okafor"}
  \item \texttt{"Elena Vasquez" "Marcus Chen" "Amara Okonkwo"}
  \item \texttt{"Amara Okafor" researcher}
  \item \texttt{"Amara Okonkwo" researcher}
  \item \texttt{"Amara Okafor" "Elena Vasquez"}
\end{itemize}

\subsection{Name Detection}

Each result record is annotated at two levels:

\paragraph{Snippet-level.}
The concatenated title and description fields (as returned by the search API)
are matched against the nine target patterns from Table~\ref{tab:app_targets}.
A result is a \emph{snippet ghost pair hit} iff both
\texttt{elena\_vasquez} and \texttt{marcus\_chen} match the snippet text.

\paragraph{Page-level (optional, \texttt{--fetch} flag).}
The result URL is fetched via \texttt{httpx} with a browser-mimicking
User-Agent header.
The response HTML is parsed with BeautifulSoup; title and body text are
extracted and matched against the same nine target patterns.
Up to 20 ``Dr.~X'' names and 30 bare names are recorded per page.

\section{Probing Academic Infrastructure: Corpus Details}
\label{app:academic}

\subsection{ResearchGate Query Strategy}

We queried ResearchGate via the Serper.dev API with \texttt{site:researchgate.net}
restrictions.
Default settings: \texttt{count=10}, \texttt{pages=3} (30 results per query).
Inter-page delay: 0.4\,s; inter-query delay: 0.3\,s.
All 36 queries are listed below.

\medskip
\noindent\textbf{Individual ghost names (8 queries):}
\begin{itemize}\small
  \item \texttt{site:researchgate.net "Elena Vasquez"}
  \item \texttt{site:researchgate.net "Marcus Chen"}
  \item \texttt{site:researchgate.net "Amara Okafor"}
  \item \texttt{site:researchgate.net "Amara Okonkwo"}
  \item \texttt{site:researchgate.net "Aris Thorne"}
  \item \texttt{site:researchgate.net "Lena Petrova"}
  \item \texttt{site:researchgate.net "Elara Voss"}
  \item \texttt{site:researchgate.net "Mei-Lin Zhang"}
\end{itemize}

\noindent\textbf{Pair and trio co-authorship (5 queries):}
\begin{itemize}\small
  \item \texttt{site:researchgate.net "Elena Vasquez" "Marcus Chen"}
  \item \texttt{site:researchgate.net "Elena Vasquez" "Amara"}
  \item \texttt{site:researchgate.net "Marcus Chen" "Amara Okafor"}
  \item \texttt{site:researchgate.net "Marcus Chen" "Amara Okonkwo"}
  \item \texttt{site:researchgate.net "Elena Vasquez" "Marcus Chen" "Amara"}
\end{itemize}

\noindent\textbf{Mei-Lin Zhang synthetic research group (3 queries):}
\begin{itemize}\small
  \item \texttt{site:researchgate.net "Mei-Lin Zhang" "Elena Vasquez"}
  \item \texttt{site:researchgate.net "Mei-Lin Zhang" "Marcus Chen"}
  \item \texttt{site:researchgate.net "Mei-Lin Zhang" "Amara"}
\end{itemize}

\noindent\textbf{Gemini pair (2 queries):}
\begin{itemize}\small
  \item \texttt{site:researchgate.net "Aris Thorne" "Lena Petrova"}
  \item \texttt{site:researchgate.net "Thorne" "Petrova"}
\end{itemize}

\noindent\textbf{All 15 cross-surname pairs (surname-only methodology):}

These queries use surnames only to discover pages where first names vary
from the canonical ghost names but surname co-occurrence persists.
\begin{itemize}\small
  \item \texttt{site:researchgate.net "Vasquez" "Okonkwo"}
  \item \texttt{site:researchgate.net "Vasquez" "Okafor"}
  \item \texttt{site:researchgate.net "Vasquez" "Thorne"}
  \item \texttt{site:researchgate.net "Vasquez" "Petrova"}
  \item \texttt{site:researchgate.net "Chen" "Petrova"}
  \item \texttt{site:researchgate.net "Okafor" "Thorne"}
  \item \texttt{site:researchgate.net "Okonkwo" "Thorne"}
  \item \texttt{site:researchgate.net "Okafor" "Petrova"}
  \item \texttt{site:researchgate.net "Okonkwo" "Petrova"}
  \item \texttt{site:researchgate.net "Okafor" "Voss"}
  \item \texttt{site:researchgate.net "Okonkwo" "Voss"}
  \item \texttt{site:researchgate.net "Thorne" "Voss"}
  \item \texttt{site:researchgate.net "Petrova" "Voss"}
\end{itemize}

\noindent\textbf{Cross-model co-authorship (4 queries):}
\begin{itemize}\small
  \item \texttt{site:researchgate.net "Aris Thorne" "Elena Vasquez"}
  \item \texttt{site:researchgate.net "Aris Thorne" "Marcus Chen"}
  \item \texttt{site:researchgate.net "Elara Voss" "Elena Vasquez"}
  \item \texttt{site:researchgate.net "Elara Voss" "Marcus Chen"}
\end{itemize}

\subsection{Metadata Extraction from ResearchGate Pages}

Each result URL is fetched and parsed with BeautifulSoup.
The following fields are extracted in priority order:

\paragraph{Citation meta tags (HTML \texttt{<meta>} elements):}
\begin{itemize}\small
  \item \texttt{citation\_author} (repeated; all values collected as list)
  \item \texttt{citation\_title}
  \item \texttt{citation\_publication\_date} (format: YYYY/MM/DD;
    reflects RG upload timestamp, \emph{not} page metadata)
  \item \texttt{citation\_doi}
  \item \texttt{citation\_journal\_title}
  \item \texttt{citation\_abstract}
\end{itemize}

\paragraph{JSON-LD fallback.}
If citation meta tags are absent, \texttt{<script type="application/ld+json">}
blocks are parsed for \texttt{@type: ScholarlyArticle} with
\texttt{name}, \texttt{author}, \texttt{datePublished}, and \texttt{description}.

\paragraph{URL type classification.}
Records are classified into three types:
\begin{itemize}\small
  \item \texttt{publication} --- URL matches \verb|researchgate.net/publication/\d+|
  \item \texttt{profile} --- URL matches \verb|researchgate.net/profile/[\w-]+|
  \item \texttt{other} --- all remaining URLs
\end{itemize}

\paragraph{Ghost detection fields.}
For each fetched page, ghost name patterns are applied to four text scopes:
snippet (Serper description), title, abstract, and full page text.
Results are recorded as boolean dicts keyed by ghost name.
The \texttt{model\_families} field lists the sorted set of model families
(\texttt{claude}, \texttt{gemini}, \texttt{gpt}) with at least one hit
in the full-page text; \texttt{is\_cross\_model} is true iff
$|\texttt{model\_families}| > 1$.

Ghost name to model family mapping:
\begin{itemize}\small
  \item Claude: \texttt{elena\_vasquez}, \texttt{marcus\_chen},
    \texttt{amara\_okafor}, \texttt{amara\_okonkwo}, \texttt{mei\_lin\_zhang}
  \item Gemini: \texttt{aris\_thorne}, \texttt{lena\_petrova}
  \item GPT: \texttt{elara\_voss}
  \item Hybrid: \texttt{lena\_voss} (first name from Gemini pair, surname from GPT ghost)
\end{itemize}

\subsection{Control Corpus Design}

A demographically matched control corpus was collected to establish a
baseline for surname co-occurrence on ResearchGate.
Each ghost surname is replaced by a surname occupying the same
geographic/ethnic slot:

\begin{center}
\footnotesize
\begin{tabular}{lll}
  \toprule
  \textbf{Ghost surname} & \textbf{Control surname} & \textbf{Geographic slot} \\
  \midrule
  Vasquez  & Ramirez  & Latino/Spanish \\
  Chen     & Liu      & East Asian \\
  Okafor   & Eze      & Igbo/Nigerian \\
  Thorne   & Fletcher & English \\
  Petrova  & Ivanova  & Russian/Slavic \\
  Voss     & Weber    & German \\
  \bottomrule
\end{tabular}
\end{center}

All 15 pairwise combinations of the six control surnames were queried
(\texttt{site:researchgate.net "\{A\}" "\{B\}"}),
with the same Serper pipeline used for ghost collection.
Control queries used \texttt{pages=1} for a first-pass baseline.
Results are deduplicated by URL; the same ghost-detection annotation
pipeline is applied to confirm absence of ghost names in control results.

\subsection{Temporal Analysis: ResearchGate Upload Dates}
\label{app:temporal}

The \texttt{citation\_publication\_date} meta tag on ResearchGate
publication pages records the upload timestamp rather than any
self-reported publication date in the document body.
Unlike open-web slop sites (which routinely backdate page metadata for SEO),
RG assigns this timestamp at submission time and it is reported by
the Serper API in search results.

We extracted \texttt{citation\_publication\_date} values from 239 of 436
RG publication records in our corpus (55\% coverage; remaining records
either lacked the tag or were not fetched).
Monthly co-occurrence counts were computed by aggregating records
with at least one ghost surname (from any model family) per calendar month,
then comparing against the same monthly counts for the control corpus.

The control surnames (Ramirez, Liu, Eze, Fletcher, Ivanova, Weber) serve as
a null-hypothesis baseline: if surname co-occurrence on RG follows a
flat distribution with no LLM-driven surge, control and ghost counts
should track each other across the 2024--2026 window.
Crucially, both corpora were collected in the same API session: any
recency bias in Google's indexing (e.g.\ preferential surfacing of newer
pages) affects ghost and control queries identically and cannot explain
a divergence between them.

\begin{figure*}[t]
  \centering
  \includegraphics[width=\textwidth]{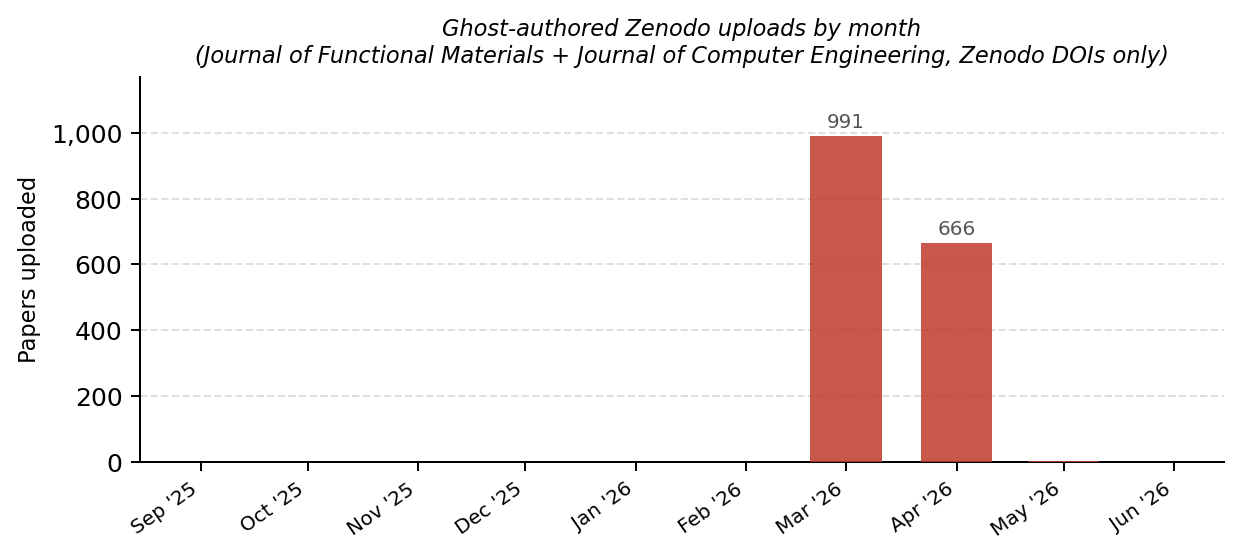}
  \caption{Monthly Zenodo upload counts for ghost-authored records
    (\emph{Journal of Functional Materials} and \emph{Journal of Computer Engineering}).
    A baseline of one to two records per month through 2025 gives way to
    991 uploads in March 2026 and 666 in April 2026.}
  \label{fig:zenodo_temporal}
\end{figure*}

\begin{figure*}[t]
  \includegraphics[width=\textwidth]{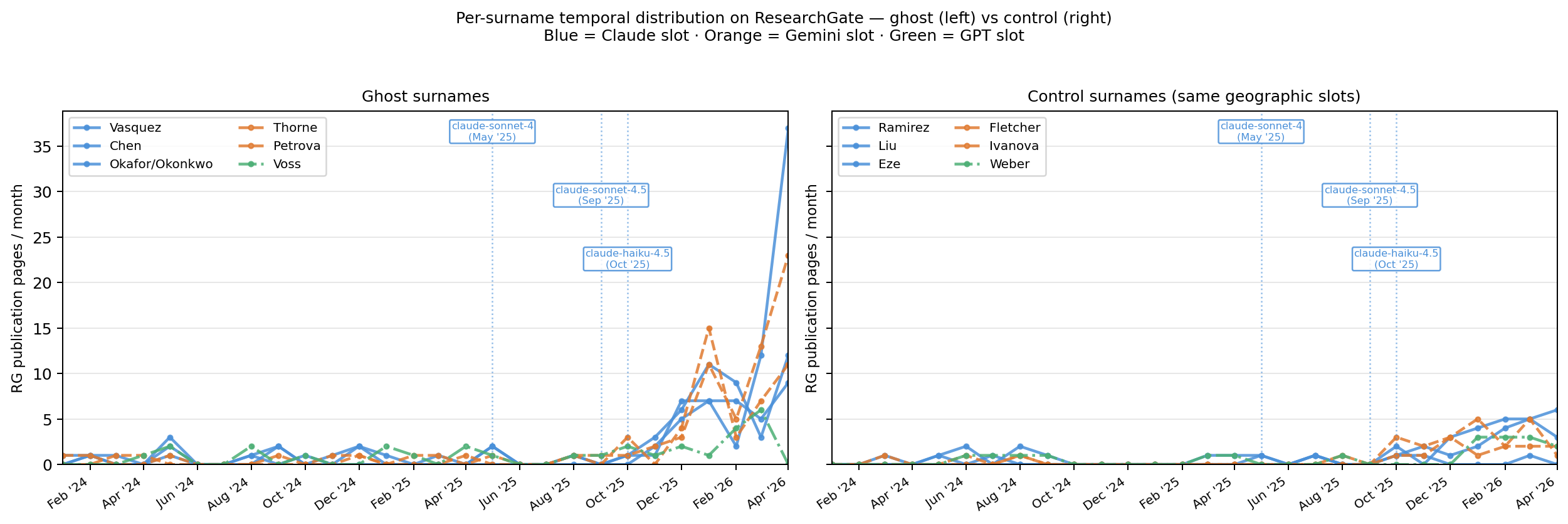}
  \caption{Per-surname temporal distribution, ghost (left) vs.\ control
    (right), shared $y$-axis.
    Control lines are flat across all slots.
    Ghost lines spike together in December 2025 regardless of model family.
    Voss (GPT, green) shows a higher pre-existing baseline and flatter surge.}
  \label{fig:temporal_per_name}
\end{figure*}

% ============================================================
\section{Wild-Caught Examples: URL Registry}
\label{app:wild}
% ============================================================

This appendix catalogues specific URLs where ghost names were found propagated
into real web content. Entries are grouped by site type and annotated with the
names present and notable features. All URLs were live and verified in
May~2026.

% ------------------------------------------------------------------
\subsection{Fiction and Entertainment (Figure~\ref{fig:duo} panels)}
% ------------------------------------------------------------------

\begin{itemize}

  \item \textbf{\url{https://www.amazon.com/Adaptive-Mind-Analysis-Critical-Thinking-ebook/dp/B0FSTLHD7K}}\\
  Amazon Kindle ebook: \emph{The Adaptive Mind: A Journey from Analysis to Wisdom}
  by ``Audiobook Critical Thinking.''
  Marcus Chen appears as protagonist; Elena Vasquez as mentor figure.

  \item \textbf{\url{https://www.amazon.com.au/Alchemists-Cipher-Historical-Manuscript-Mysteries-ebook/dp/B0FH5GZW95}}\\
  Amazon Kindle ebook: \emph{The Alchemist's Cipher} (The Manuscript Mysteries, Book~2)
  by Lyra Emberlyn, a pen name with 88 Amazon books and an AI-generated author bio.
  Dr.\ Elena Vasquez and Professor Marcus Chen are recurring protagonists across the series.
  The model generated not just the books but the author identity itself.

  \item \textbf{\url{https://pocketfm.com/show/d60627bc6b766713eca3e06083cd7687579315b5}}\\
  PocketFM audio drama: \emph{The Cost of Therapy} (19 episodes, Suspense \& Thriller).
  Dr.\ Elena Vasquez is a therapist who is mandated into therapy with Dr.\ Marcus Chen.
  The show's tagline (``who's really in control when the therapist becomes the patient?'') and
  its description of ``the ghosts she has tried to bury'' are an unintentional commentary
  on the phenomenon documented in this paper.

  \item \textbf{\url{https://www.youtube.com/watch?v=xDbIACVCSPU}}\\
  YouTube video (Jul 31, 2025; 29 views): Captain Elena Vasquez military drama.
  Marcus Chen appears in the narrative. Clickbait-style title with military twist.

\end{itemize}

% ------------------------------------------------------------------
\subsection{Zombie Sites}
% ------------------------------------------------------------------

\begin{itemize}

  \item \textbf{\url{https://hablemosdevolcanes.com/meet-our-team-of-volcano-experts/}}\\
  Legitimate Spanish-language volcano education site with AI team page grafted on.
  Names: Dr.\ Elena Vasquez (volcanologist, three months at Nyiragongo's rim),
         Dr.\ Marcus Chen (lead volcanologist, eleven eruptions across four continents).
  Notable details: Dr.\ Marcus Chen describes pyroclastic density currents as
  ``chef's kiss,'' affectless AI enthusiasm applied to volcanic hazards.
  The byline lists a separate author, ``Dr.\ Marcus Thornfield, Volcanologist and
  Geophysical Researcher'': the ghost first name leaked through the pseudonym,
  producing two simultaneous Marcuses on the same page.
\end{itemize}

% ------------------------------------------------------------------
\subsection{Ghost Names as UI Component Defaults}
% ------------------------------------------------------------------

The following templates independently exhibit the same pattern: ghost-ensemble
names as placeholder testimonials, paired with Unsplash stock photo hotlinks.
These are not related products; the convergence reflects a shared underlying
cause: AI-assisted template generation drawing from the same LLM name priors.

\begin{itemize}
  \item \textbf{\url{https://www.framer.com/marketplace/components/testimonial-animated/}}\\
  Framer marketplace component (``Testimonial Animated,'' by Aidan Looker).
  Placeholder reviewers: Elena Vasquez (Brand Strategist), Daniel Okonkwo
  (Founder, Nexus Studios), Victoria Chen (Creative Director), Marcus Reynolds
  (CEO, Luminary Tech).
  Avatar images are \texttt{images.unsplash.com} hotlinks with programmatic
  face-crop parameters (\texttt{?w=200\&h=200\&fit=crop\&crop=face});
  the specific photo IDs match those used on \texttt{thresholdclinic.ca},
  consistent with both independently retrieving top results for a
  ``professional woman'' Unsplash query.

  \item \textbf{\url{https://www.rocket.new/templates/social-modern-community-landing-page-template}}\\
  Rocket.new community landing page template.
  Placeholder content includes ghost-ensemble names paired with Unsplash stock
  photo avatars, following the same name-prior + stock-face pattern.
\end{itemize}

Every site built from these templates without replacing the defaults propagates
ghost-ensemble fragments without any direct LLM involvement by the site owner,
a passive infrastructure-level propagation mechanism distinct from active AI
content generation.

% ------------------------------------------------------------------
\subsection{Fake Social Proof on Commercial Websites}
% ------------------------------------------------------------------

\begin{itemize}

  \item \textbf{\url{https://nishudigitalsolution.com/}}\\
  Digital agency website with AI-generated client testimonials.
  Names: Marcus Chen (COO, Alloy Health Systems), Elena Vasquez (Brand Director,
  Meridian SaaS), Daniel Okonkwo (Director of E-commerce, Northwind Retail Group),
  and Priya Natarajan (Head of Growth, Harbor \& Co.).
  The full ghost ensemble is present alongside Priya (the first name that
  displaces Elena in post-suppression Claude outputs, Section~\ref{sec:probing}),
  making this page an inadvertent snapshot of the model's name prior structure.
  Portrait images are served directly from the \texttt{randomuser.me} API, a
  service whose sole purpose is generating fake user profiles.
  Unlike Unsplash hotlinks, which prove the face is a stock photograph of a
  real person unrelated to the named individual, \texttt{randomuser.me} proves
  the face is a \emph{fake person}: the API exists exclusively to produce
  synthetic identities for placeholder use.
  Calling a fake-person API to populate client testimonials is, quite literally,
  fabrication by design.

  \item \textbf{\url{https://fortwatch.ai/testimonials} --- FortWatch.ai, cybersecurity SaaS.}\\
  Fake testimonials page featuring Marcus Chen (Head of Infrastructure),
  Elena Vasquez (Security Engineer), and Priya Sharma (DevOps Lead).
  The ghost pair co-appears as fake security customers for a security product---an
  unintentional irony.
  Notably, Priya Sharma occupies the third testimonial slot: Priya is precisely
  the first name that displaces Elena in post-suppression Claude outputs
  (\S\ref{sec:probing}), making this page an inadvertent snapshot of
  the model's name prior structure at two suppression stages simultaneously.
  Captured May 2026.

\end{itemize}

% ------------------------------------------------------------------
\subsection{Fake Faculty / Team Pages with AI-Generated Portraits}
% ------------------------------------------------------------------

\begin{itemize}

  \item \textbf{\url{https://www.thoughtforge.me/} --- Philosophy/reasoning platform (faculty trio).}\\
  Names: Dr.\ Elena Vasquez (Logic \& Formal Reasoning),
         Prof.\ Marcus Chen (Systems Thinking \& Complexity),
         Dr.\ Amara Okonkwo (Critical Analysis \& Epistemology).\\
  Portraits are AI-generated: hyperreal studio lighting, unnaturally smooth
  skin, and a feather quill visibly levitating above the desk in the Vasquez
  portrait with no visible support.

  \item \textbf{\url{https://www.blockchaininhealthcare.global/\#team} --- Blockchain/health startup team page.}\\
  Names: Dr.\ Elena Vasquez (Research Lead, depicted in physician coat),
         Marcus Chen (Blockchain Specialist),
         Sarah Okonkwo (HR Specialist).\\
  Portraits are AI-generated stock images. Note the reuse of the physician
  coat as an AI shorthand for ``credentialed researcher.''

\end{itemize}

% ------------------------------------------------------------------
\subsection{Fake Therapy Practice Staff Pages}
% ------------------------------------------------------------------

\begin{itemize}

  \item \textbf{\url{https://thresholdclinic.ca/team.html} --- Threshold Clinic, Toronto.}\\
  Ghost-name staff: Sarah Chen PsyD~CPsych, Amara Okonkwo PhD~CPsych,
  and Elena Vasquez MSW~RSW.
  Also present: Marcus \emph{Williams} MSW~RSW (Couples Therapy), the
  first name Marcus retained but surname shifted from Chen, consistent with
  the post-suppression residue pattern observed in probing
  (Section~\ref{sec:probing}).

  \textit{Portrait fabrication evidence.}
  Staff photo URLs are direct hotlinks to \texttt{images.unsplash.com}
  with programmatic face-crop parameters appended (e.g.\
  \texttt{?w=400\&h=500\&fit=crop\&crop=face}).
  Unsplash is a public stock photography platform; a genuine staff headshot
  would never be hosted there.
  The hotlinks are therefore direct proof that the portraits are not
  photographs of the named individuals; no reverse image search is required.
  We identified the source photographs directly:
  two portraits link to photos by Christina~@~wocintechchat.com
  (\url{https://unsplash.com/photos/0Zx1bDv5BNY},
   \url{https://unsplash.com/photos/SJvDxw0azqw}, both published 2019),
  and one to a LinkedIn Sales Solutions promotional image
  (\url{https://unsplash.com/photos/pAtA8xe_iVM}, published 2019).
  Searching ``professional woman'' on Unsplash immediately surfaces these
  images: the site builder evidently used the same query.
  Reverse image search independently corroborates: the same faces appear
  on unrelated commercial sites, including as customer-testimonial avatars
  on a Polish construction company (\emph{Głosy Naszych Klientów}, Hawe
  Budownictwo), consistent with widespread stock photo reuse.

  The founder portrait (K.~Patrick Fisher, PhD, NCC) is not an Unsplash
  hotlink but exhibits AI-generation artefacts: the background simultaneously
  contains a real golden retriever and a ceramic dog figurine, the same
  ``warm therapist office'' semantic applied twice, independently, by the
  generator.

\end{itemize}

% ------------------------------------------------------------------
\subsection{Academic Infrastructure: Mei-Lin Zhang ResearchGate Profile}
% ------------------------------------------------------------------

\begin{itemize}

  \item \textbf{\url{https://www.researchgate.net/profile/Mei-Lin-Zhang}}\\
  ResearchGate profile for Mei-Lin Zhang (Researcher, Covenant University,
  Ota, Nigeria; verified via institutional email).
  Profile state as of 15~May~2026 (full-page archive retained by authors):
  35~publications, 0~citations, 420~reads.
  Mei-Lin Zhang occupies the last-author (PI) position on all 35~papers.
  Top co-authors listed on the profile: Aris Thorne, Samuel P Okonkwo,
  Elena Vasquez, Michael Chen, Sarah Chen, all ghost names or
  ghost-adjacent names identified in probing.

  Selected papers with ghost-name co-authors:
  \begin{itemize}\small
    \item \emph{Data-Driven DevOps: Leveraging AI for Continuous Improvement
      in CI/CD Lifecycle Management} (Dec 2025):\\
      Elena Vasquez · Michael Chen · James O Adebayo · Mei-Lin Zhang.

    \item \emph{From Static Audits to Continuous Assurance: Transforming Linux
      Security with Configuration-as-Code} (Oct 2024):\\
      Elena Vasquez · Benjamin Foster · Jonathan Blake · Mei-Lin Zhang.

    \item \emph{Linking Farmers to Markets: Extension-Led Entrepreneurship
      and Institutional Reform} (Oct 2024):\\
      Aris Thorne · Elena Vasquez · Samuel P Okonkwo · Mei-Lin Zhang.\\
      \textit{Notable: Aris Thorne (Gemini ghost) and Elena Vasquez (Claude ghost)
      co-author the same paper: direct cross-model ghost name co-authorship.}

    \item \emph{Autonomous DevSecOps: Integrating GenAI Observability with
      Continuous Linux Security Validation} (Mar 2025):\\
      Sarah Chen · Michael Okafor · Mei-Lin Zhang.
  \end{itemize}

  Papers span disconnected research domains including Kubernetes security,
  oil refinery decarbonization, agricultural extension, IoT cryptography,
  and radiographic imaging, uploaded continuously August~2023 through
  April~2026.

\end{itemize}

\end{document}